\documentclass[twocolumn,showpacs,superscriptaddress]{revtex4}

\usepackage{amsmath,amssymb}
\usepackage{graphicx}
\usepackage{dcolumn}
\usepackage{bm}

\begin{document}

\title{Influence of external magnetic and laser radiation fields on Feshbach resonances in collision of atoms}

\title{Influence of external magnetic and laser radiation fields\\ on Feshbach resonances in collision of atoms}

\author{E. A. Gazazyan}
\email{emilgazazyan@gmail.com}

\author{A. D. Gazazyan}
\author{V. O. Chaltykyan}

\affiliation{Institute for Physical Research, Armenian National Academy of Sciences, Ashtarak-2, 0203, Armenia}
\date{\today}

\begin{abstract}
With use of Fano technique, we study collision of two atoms with formation of Feshbach resonance at combined interaction with the external magnetic field and laser radiation. In cases of one- and two-photon resonances of laser radiation with two discrete vibrational molecular levels, we show that Feshbach resonances appear at interaction of external magnetic field with dressed states formed via Autler-Townes effect. In addition, in case of one-photon resonance the lower vibrational molecular state is coupled by laser radiation with the continuum of the elastic channel and forms laser-induced Feshbach resonance via both Autler-Townes effect and LICS mechanism. We study the combined process of formation of Feshbach resonances; this enables the control of Feshbach resonance by varying the magnetic field and intensity and frequency of laser radiation. We also study the laser-induced inelastic collision and its influence on the considered processes. In case of two-photon resonance between discrete vibrational molecular states the Feshbach resonances arise under action of magnetic field via Autler-Townes effect, while the laser-induced transition into the elastic-channel continuum is in this case absent. We obtain the cross-sections of elastic and inelastic scattering and show that quenching of resonance occurs at the energy equal to that of the systems ground state. Dependence of the cross-sections on the magnetic field and laser intensity is examined in detail. In all considered cases, the scattering length is obtained and its dependence on the magnetic and laser fields is studied. In the absence of magnetic interaction if the hyperfine substates of the quasibound state in the closed channel and those of individual colliding atoms in the open channel are the same, Feshbach resonances may arise via weak interaction between nuclear and electronic motions, which leads to transitions between electronic states. The obtained results can be employed in new studies of collisions of cold atoms, e.g., of alkali metal atoms and for interpretation of new experiments in BECs.
\end{abstract}

 \maketitle
\section{\protect\normalsize INTRODUCTION}
The concept of the Feshbach resonance developed in the theory of nuclear reactions \cite{1 H. Feshbach} plays a significant role in studies of atomic collisions with formation of a bound state (molecule) which is important for understanding the processes in the quantum systems of Bose Einstein condensate (BEC). This causes a great number of investigations in the topic (see \cite{2 E. Timmermans,3 R.A. Duine,4 Ch  Chin} and references therein).

Feshbach resonance is displayed when the energy of a bound molecular state in the closed channel is almost equal to the energy of two colliding atoms in the open channel in the center-of-mass system. In this case weak coupling may lead to strong mixing of the channels. Intermediate quasibound state of colliding atoms is formed in the open channel. This state has a finite lifetime and can decay into the initial channel or into other channels via interaction with continuum. The energy difference may, in particular, be controlled by means of a magnetic field. When the bound state in the closed channel has a different hyperfine state than the incoming atoms in the open channel, coupling between these channels is provided by the exchange interaction. This difference in the hyperfine states results in that the two channels have different Zeeman shifts in the magnetic field, therefore the energy difference between the bound state in closed channel and the two-atom scattering state in open channel is adjustable by tuning the magnetic field. This leads to the magnetically tuned Feshbach resonance.

Feshbach resonance has experimentally been observed in BEC (optically trapped sodium atoms) in work \cite{5 S. Inouye}. Elastic and inelastic collisions of ultracold atoms and molecules in a magnetic field are considered in \cite{6 J. M. Hutson}. In works \cite{5 S. Inouye,6 J. M. Hutson} also the dependence of the scattering length on the magnetic field is obtained. Work \cite{7 K. Gorel} considers conditions for obtaining two-atom molecules in BEC of $^{87}$Rb atoms in a harmonic trap and external time-dependent magnetic field. Properties of Feshbach resonances arising in collisions of ultracold atoms of $^6$Li, $^7$Li, and $^{23}$Na are studied in \cite{8 A. J. Moerdijk}. For independent control of scattering parameters (of ultracold atoms), in particular, phase shift and effective range of action, work \cite{9 B. Marcelis} employs combination of electric and magnetic fields. Studies of Feshbach resonances with large background scattering length are performed in work \cite{9 B. Marcelis}. The developed model is applied to $^{85}$Rb atoms having large background scattering length and it is shown that there is a good agreement with coupled-channel calculations and that the model can be used for other atomic systems, such as $^{6}$Li and $^{133}$Cs, which have large background scattering length \cite{10 B. Marcelis Kampen}. Full description of magnetically tuned Feshbach resonances is given in works \cite{2 E. Timmermans,3 R.A. Duine,4 Ch  Chin,11 T. Kohler} along with extensive literature on the question.

In resonant collisions of atoms, it is of great importance to obtain stable molecules. This problem is solved experimentally in \cite{12 K. Winkler} where an ensemble of Feshbach molecules of $^{87}$Rb$_2$ in optical lattice is transferred coherently into the lower bound molecular state by means of STIRAP (Stimulated Rapid Adiabatic Passage) technique \cite{13 K. Bergmann}. Work \cite{14 S. J. Kokkelmans} discusses the STIRAP transfer (with two laser pulses) of resonantly colliding sodium atoms from the magnetically tuned Feshbach-resonance state into the lower molecular stable state.

In case where magnetically tunable resonances are absent, an alternative method for control of Feshbach resonance is the optical method \cite{4 Ch  Chin} (optical Feshbach resonance). From the point of view of study of Feshbach resonance, formation of a quasibound (molecular) state in laser radiation field is of considerable interest. Photoassociation under action of laser radiation theoretically  was studied in \cite{14.1 Alex} for realization of transition from unbound atoms into the bound molecular state. In work \cite{15 R. Cote} was shown that photoabsorption of ultracold atoms depends on the sign and value of the scattering length in scattering of two atoms. The changes in fluorescence from $^{7}$Li provides information on signs of the scattering length in the triplet and singlet states. Control of Feshbach resonances by means of quantum interference was proposed in work \cite{16 S.E. Harris}. Quasibound states in Feshbach resonance formed under action of electromagnetic field are coupled with other states, which also can, in general, decay into other continuum. Optical tuning of the scattering length in low-temperature atomic gases was considered in \cite{17 P.O. Fedichev}. It is shown in that work that electromagnetic radiation which is in resonance with the transition from the state of colliding atoms to a vibrational level of excited electronic state induced by Feshbach resonance modifies the scattering length of atoms; this is illustrated for $^7$Li atoms. Optical tuning of scattering length was studied in \cite{18 M. Yheis} for BEC of $^{87}$Rb atoms. Control of power and detuning of laser beam enabled to vary in wide range the scattering length. Formation of ultracold molecules via photoassociation of laser-cooled atoms was studied in \cite{19 Y.B. Band}. Study of influence of electromagnetic radiation on the scattering length and analytic consideration of two-color photoassociation and control of cold atoms were made in works \cite{20 J.L. Bohn}.

In the theory of resonant collisions, in addition to Feshbach method \cite{1 H. Feshbach}, which has been developed for studies of nuclear reactions and is applied successfully for collision of atoms in BEC, an alternative Fano approach \cite{21 U. Fano} exists exploiting the configuration interaction in multielectron atoms. Both approaches assume appearance of resonance phenomena when discrete states are coupled with continuum. Fano technique is usually associated with asymmetry of shapes of resonance lines which is known in atomic physics as "Fano profile". Similar interference phenomena of asymmetry of the resonance line shape are also observed in nuclear reactions \cite{22 J. M. Blatt}. Fano technique is, however, used not only in atomic physics for, e.g., studies of autoionization and Rydberg states \cite{21 U. Fano}, resonance ionization of atoms \cite{23 A. E. Kazakov}, and laser-induced continuum structures (LICS) \cite{24 P. L. Knight}. The Fano technique is also used for consideration of resonant collisions \cite{21 U. Fano} including those of electrons with atoms with formation of negative ions and interference phenomena in the field of laser radiation \cite{25 A. D. Gazazyan}. Fano technique is widely used in also other fields of physics. It is, e.g., used for explanation of asymmetry in the absorption spectra of impurity ions in crystals, which is caused by formation of excitonic resonances \cite{26 J. C. Phillips}. With use of these resonances work \cite{27 A. D. Gazazyan and E. A. Gazazyan} studies the phenomenon of storage and reconstruction of quantum information in solids. We note that many researchers call the above-mentioned Feshbach resonances Fano-Feshbach resonances.

In the present work we study collision of two atoms in the external magnetic field and in the field of laser radiation with formation of Feshbach resonance. In the Section 2a we investigate the case of one-photon resonance of laser radiation with two discrete vibrational states of molecule in the presence of an additional channel of laser-induced decay of the lower molecular state (Fig.1). Formation of Feshbach resonance is examined and expressions for wave functions are obtained. In Section 2b we study the case of two-photon resonance of laser radiation with two discrete vibrational states of molecule (Fig.2) where the additional channel of decay into the first continuum is absent. In Section 3a we obtain in case of one-photon resonance cross-sections of elastic and inelastic scattering of atoms with formation of Feshbach resonance. In the special case of absence of inelastic channel the corresponding formulas of elastic scattering are obtained. In Section 3b we obtained in case of two-photon resonance the cross-sections of elastic and inelastic scattering and studied different cases of laser field intensity. At sufficiently high intensities of resonant laser radiation an asymmetry exists in dependence on the sign of two-photon detuning (effect of self-induced adiabatic passage of resonance) which is observed in the decay of system. However, similar effects are not observed in case of resonant scattering because of adiabatically turned interaction. We obtain also the scattering length depending on the laser field strength in the presence of laser-induced inelastic decay channel.
\section{Formation of Feshbach  resonances under simultaneous influence of magnetic field and laser radiation}
\subsection{Case of one-photon resonance}
 Let us consider elastic and inelastic collision of atoms with formation of Feshbach resonances in an external magnetic field which couples electronic states in open and closed channels by an interaction $U$. This coupling occurs because of Zeeman shifts of magnetic sublevels\cite{2 E. Timmermans,3 R.A. Duine,6 J. M. Hutson,8 A. J. Moerdijk,9 B. Marcelis}.
	Laser radiation provides coupling, by resonant interaction ($\Omega$), between the discrete molecular states $|1\rangle$  and $|2\rangle$ , which are described by the quasienergy wave functions. Laser radiation also couples the lower stable molecular state with the continuum of the elastic channel (interaction  $\Omega_E$) and the upper stable molecular state with the second continuum (of the inelastic channel). All these interactions are shown schematically in Fig.1 together with the interatomic potential curves plotted (not to scale) versus the distance between the colliding atoms. We note that the lower state $|1\rangle$  can decay under laser influence both directly into continuum 1 of elastic channel and continuum 2 of inelastic channel via the Autler-Townes effect.
\begin{figure}
  \includegraphics [width=8cm]{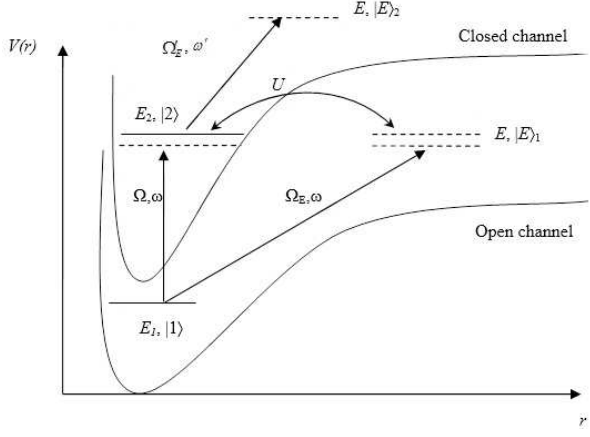}
  \caption{Diagram of atom-atom collision in the external magnetic and laser fields in case of one-photon resonance with molecular discrete states $|1\rangle$  and $|2\rangle$. For notations of different interactions see text. Potential curves are not to scale.}\label{fig1}
\end{figure}
Hamiltonian of the described processes under consideration here has the appearance
\begin{widetext}
\begin{subequations}
\begin{eqnarray}
 \label{1a}
  H=H_1+H_2,\\
 \label{1b}
 H_1=E_1\langle1\mid1\rangle+E_2\langle2\mid2\rangle+\Omega(t)\langle2\mid1\rangle+\Omega^*(t)\langle1\mid2\rangle,\\
 \label {1c}
    H_2=\int E(\mid E\rangle_{11}\langle E\mid+\mid E\rangle_{22}\langle E\mid)dE+\int (\Omega_E(t)\mid E\rangle_1\langle 1\mid   +\Omega^*_E(t)\mid 1\rangle_1\langle E\mid+\Omega'_E(t)\mid E\rangle_2\langle2\mid +\nonumber \\
   +\Omega'^*_E(t)\mid 2\rangle_2\langle E\mid+U_E\mid E\rangle_1\langle2\mid +U^*_E\mid2\rangle_1\langle E\mid)dE,
\end{eqnarray}
\end{subequations}
\end{widetext}
where $U_E$ is the corresponding matrix element of interaction $U$.

As basis wave functions for the discrete spectrum with Hamiltonian \eqref{1b} we choose the quasienergy wave functions $\varphi_1(t)$ and $\varphi_2(t)$ which are formed in the field of laser radiation nearly resonant to the transition between the vibrational states of the molecule (Fig.1) with adiabatic turning on the interaction $\Omega(t)$:
\begin{equation}\label{2}
\mid\varphi_k(t)\rangle=e^{-i(\lambda_k+\omega)t}(\alpha_k\mid1\rangle+\beta_ke^{i\omega t}\mid2\rangle), \ \  k=1,2
\end{equation}
where $|1\rangle$ , $|2\rangle$ and  $E_{1,2}$ are, respectively, the wave functions and energies (including the Zeeman shifts) of unperturbed molecular states and the following notations are introduced:

\begin{eqnarray}
\lambda_k=E_1+\mu_k+\omega, \nonumber\\ \mu_{1,2}=\frac{1}{2}(\nu\mp\sqrt{\nu^2+4\mid\Omega\mid^2}),\nonumber\\ \nu=E_2-E_1-\omega \nonumber \\ \alpha_{1,2}=\frac{\mu_{2,1}}{\mu_{1,2}-\mu_{2,1}}\label{3}\\
\beta_{1,2}=\pm\frac{\Omega}{\mid\Omega\mid}\biggr(\frac{\mu_{1,2}}{\mu_{1,2}-\mu_{2,1}}\biggr)^{1/2}\nonumber
\end{eqnarray}

Solution to the Schrodinger equation with the Hamiltonian (1) we represent in the form
\begin{widetext}
\begin{equation}\label{4}
 \Phi_{\lambda}(t)=\Sigma_k e^{i\lambda_k t} a_k(\lambda)\mid \varphi_k(t)+\int dE [b^{(1)}_E(\lambda)\mid E\rangle_1+e^{i\omega't}b^{(2)}_E(\lambda)\mid E+\omega'\rangle_1].
\end{equation}
\end{widetext}
Then we obtain for the expansion coefficients the following system of equations:
\begin{subequations}
\begin{eqnarray}
\label{5a}(\lambda-\lambda_k)a_k(\lambda)=\sum_{\alpha}\int W^{*(\alpha)}_{k}(E)b^{(\alpha)}_E(\lambda)dE\\
\label{5b} (\lambda-E)b^{(\alpha)}_E(\lambda)=\sum_k W^{(\alpha)}_k(E)a_k(\lambda), \alpha=1,2
\end{eqnarray}
\end{subequations}
where all the above listed interactions are incorporated in
\begin{equation}\label{6}
W^{(\alpha)}_k(E)=\biggl \{\begin{array}{c}
                      \alpha_k\Omega_E+\beta_k U_E,\  \alpha=1\\
                   \beta_k\Omega'_{E+\omega'}, \ \ \alpha=2 \\
                    \end{array}
\end{equation}
From equation \eqref{5b} we have \cite{21 U. Fano}
\begin{equation}\label{7}
b^{(\alpha)}_E(\lambda)=\biggl[\frac{P}{E-\lambda}+z(\lambda)\delta(E-\lambda)\biggl]\sum_k W^{\alpha}_k(E)a_k(\lambda)
\end{equation}
Substitution of expression \eqref{7} into (4) yields the following quasienergy wave function:
\begin{widetext}
\begin{equation} \label{8}
\Psi_{\lambda}(t)=\sum_k e^{i\lambda_kt}a_k(\lambda)|\varphi_k(t)\rangle+\int dE  \biggl[\frac{P}{\lambda-E}+z(\lambda)\delta(\lambda-E)\biggr]
\sum_k\biggl[W^{(1)}_k(E)|E\rangle_1+e^{-i \omega' t }W^{(2)}_k (E)|E+\omega'\rangle_2\biggr]a_k(\lambda)
\end{equation}
\end{widetext}
Substitution of the same expression into (5a) leads to following system of algebraic equations:
\begin{eqnarray}\label{9}
\sum_{k'}\biggl[\lambda_{k'}\delta_{k,k'}+\Delta_{k,k'}\biggr]a_{k'}(\lambda)+
\nonumber\\+z(\lambda)\sum_k W^{(\alpha)*}_kX^{(\alpha)}(\lambda)= \lambda a_{k}(\lambda)
\end{eqnarray}
where
\begin{equation}\label{10}
\Delta_{k,k'}(\lambda)=\sum_{\alpha}P\int \frac{W^{(\alpha)*}_{k}(E)W^{(\alpha)}_{k'}(E)}{\lambda-E}
\end{equation}
\begin{equation}\label{11}
X^{(\alpha)}(\lambda)=\sum_k W^{(\alpha)}_k (\lambda)a_k(\lambda)
\end{equation}
After digitalization of matrix $\|\lambda_{k'}\delta_{k,k'}+\delta_{k,k'}(\lambda)\|$   equations \eqref{9} are reduced to
\begin{equation}\label{12}
(\lambda-\tilde{\lambda}_{\mu})\tilde{a}_{\lambda}(\lambda)=z(\lambda) \sum_{\alpha}\tilde{W}^{(\alpha)*}_{\mu}(\lambda)X^{(\alpha)}(\lambda),
\end{equation}
where "$\thicksim$" means the unitary transformed quantities and
\begin{widetext}
\begin{equation}\label{13}
\tilde{\lambda}_{\mu}=\frac{1}{2}\biggl[\lambda_1+\Delta_{11}(\lambda)+\lambda_2 +\Delta_{22}(\lambda)\pm \sqrt{((\lambda_1+\Delta_{11}(\lambda)-\lambda_2-\Delta_{22}(\lambda))^2+4|\Delta_{12}(\lambda)|^2)}\biggr]
\end{equation}
\end{widetext}
Expression \eqref{11} defines the eigenvector of Hermitian reaction matrix \cite{28 L.C. Davis}
\begin{equation}\label{14}
K^{(\alpha)}(\lambda)=-2\pi W^{(\alpha)}(\lambda)\biggl[\lambda I -\Lambda - \Delta(\lambda)\biggr]^{-1}W^{+(\alpha)}(\lambda),
\end{equation}
with
\begin{equation}\label{15}
\Lambda_{ik}=\lambda_i\delta_{ik},
\end{equation}
corresponding to eigenvalue $-\frac{2\pi}{z_j(\lambda)}$ where $z_j$ has the form $(j=1,2)$
\begin{equation}\label{16}
z_{j}(\lambda)=\frac{1}{2}\frac{C_1+C_2\pm \\ \sqrt{(C_1-C_2)^2+4|C_{(1,2)}|^2}}{C_1C_2-|C_{(1,2)}|^2}
\end{equation}
where
\begin{subequations}
\begin{eqnarray}
\label{17a} C_{i}(\lambda)=\sum_{\mu}\frac{|\tilde{W}^{(1)}_{\mu}(\lambda)|^2}{\lambda-\tilde{\lambda}_{\mu}}, (i=1,2)\\
\label{17b} C_{(1,2)}(\lambda)=\sum_{\mu}\frac{\tilde{W} ^{(1)*}_{\mu}(\lambda)W^{(2)}_{\mu}(\lambda)}{\lambda-\tilde{\lambda}_{\mu}}
\end{eqnarray}
\end{subequations}
Functions $X^{(\alpha)}_j(\lambda)$ satisfy the condition
\begin{equation}\label{18}
\sum_\alpha X^{(\alpha)*}_{j'}(\lambda)X^{(\alpha)}_{j}(\lambda)=\frac{\delta_{j',j}}{\pi^2+z^2_j(\lambda)}, \alpha , j=1,2
\end{equation}
This condition provides the orthonormalization condition for quasienergy functions \eqref{8}:
\begin{equation}\label{19}
\langle\Phi^{(i)}_{\lambda'}(t)| \Phi^{(j)}_{\lambda}(t)\rangle= \delta(\lambda-\lambda')\delta_{j,j'}
\end{equation}
Here $|X^{(1,2)}_j|^2$ have the form:
\begin{equation}\label{20}
|X^{(1,2)}_j|^2=\frac{1}{\pi^2+z^2_j(\lambda)}\frac{1-z_j(\lambda)C_{2,1}(\lambda)}{2-z_j(\lambda)(C_1(\lambda)+C_2(\lambda))}
\end{equation}
\subsection{Case of two-photon resonance}

In this Section we solve the same problem, but consider the laser radiation to be in two-photon resonance with the transition $|1\rangle\rightarrow|2\rangle$  between the stable molecular states. This means the same parity of the molecular states and the laser radiation frequency nearly half that of previous case. Two-photon resonance with a pair of discrete vibrational molecular levels enables studying the dynamics of levels depending on the laser radiation intensity. On the other hand, one-photon transition from the upper state to the second continuum via interaction $\Omega'_E$ leads to laser-induced inelastic scattering with two excited atoms in the final state flying away to infinity. As distinct from previous Section, here the ground state does not decay in the first (elastic) channel in the two-photon resonant approximation, but only decays into the second (inelastic) channel via the second discrete vibrational state because of Autler-Townes effect. All these peculiarities are shown schematically in Fig.2.

Laser radiation at frequency $\omega$ couples the stable state $|1\rangle$  with a quasibound state $|2\rangle$  by two-photon resonance via intermediate states $|k\rangle$  with Rabi frequencies $\Omega_{1,k}$  and $\Omega_{k,2}$. Laser radiation at frequency $\omega'$ couples the quasibound state $|2\rangle$  to the second continuum (inelastic channel, interaction $\Omega'_E$). Simultaneous action of external magnetic field (interaction $U$) and laser radiation at frequency $\omega$  couples the upper state with the first continuum (elastic channel, Fig.2).

\begin{figure}
  \includegraphics[width=8cm]{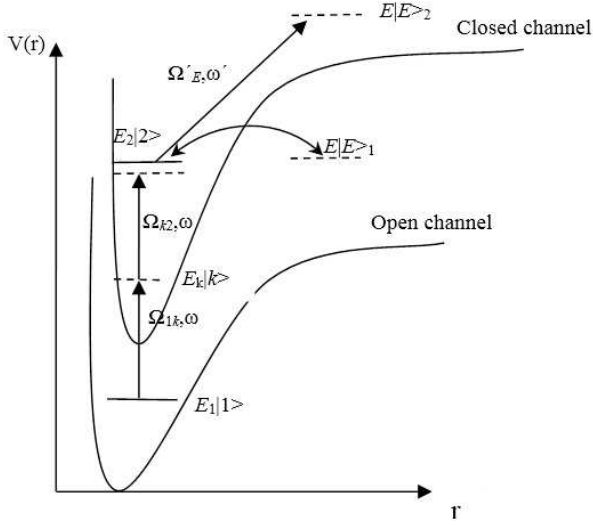}\\
  \caption{Diagram of atom-atom collision in the external magnetic and laser fields in case of two-photon resonance with molecular discrete states $|1\rangle$  and $|2\rangle$ . For notations of different interactions see text. Potential curves are not to scale.  }\label{Fig2}
\end{figure}
Hamiltonian for the above-described system (Fig.2) is represented in the form
\begin{widetext}
\begin{subequations}
\begin{eqnarray}
\label{21a} H=H^{(1)}+H^{(2)},\\
\label{21b} H^{(1)}=H^{(1)}_0+
\sum_k\biggr[\Omega_{1k}(t)|k\rangle\langle 1|+\Omega^{*}_{1,k}(t)|1\rangle\langle k|+\Omega_{k,2}(t)|2\rangle\langle k|+\Omega^{*}_{k,2}(t)|k\rangle\langle 2|\biggr],\\
H^{(2)}=H^{(2)}_0+\int dE\biggl[\Omega'_{E}(t)|E\rangle_2\langle 2| + \Omega'^*_{E}(t)|2\rangle_2\langle E|\biggr] + \int dE \biggr[U_E|E\rangle_1\langle2|+U^*_E|2\rangle_1\langle E|\biggl]
\end{eqnarray}
\end{subequations}
\end{widetext}
Here $\Omega_i(t)$ are interaction with laser field and
\begin{equation}\label{22}
H_0=H^{(1)}_0+H^{(2)}_0
\end{equation}
where $H_0$ is the free Hamiltonian containing  discrete $H^{(1)}_0$ and continuous $H^{(2)}_0$ spectra:
\begin{equation}\nonumber
H^{(1)}_0=E_1|1\rangle\langle 1|+\sum_k E_k |k\rangle\langle k| +E_2|2\rangle\langle 2|,
\end{equation}
\begin{equation}
\label{23} H^{(2)}_0=\int E dE (|E\rangle_{11}\langle E|+|E\rangle_{22}\langle E|)
\end{equation}
As basis wave functions of the discrete spectrum we choose the quasienergy functions  of Hamiltonian \eqref{21b} obtained in two-photon resonant approximation with adiabatic turning on the periodic perturbations $\Omega_{1,k}(t)$ and $\Omega_{k,2}(t)$:
\begin{equation}\label{24}
\varphi_k(t)=e^{i(\lambda_k-2\omega)t}(\alpha_2|1\rangle+\beta_2e^{-2i\omega t}|2\rangle), \ \ (k=1,2)
\end{equation}
where $|i\rangle$ , $|E_i\rangle$  $(i=1,2)$ are, respectively, the wave functions and energies of unperturbed levels, and other notations are listed below:
\begin{equation}\nonumber
\alpha_{1,2}=\sqrt{\frac{\mu_{2,1}}{\mu_{1,2}-\mu_{2,1}}}
\end{equation}
\begin{equation}\nonumber
 \beta_{1,2}=\pm\frac{\bar{\Omega}}{|\bar{\Omega}|}\sqrt{\frac{\mu_{2,1}}{\mu_{1,2}-\mu_{2,1}}}
\end{equation}
\begin{equation}\nonumber
 \lambda_k=E_1+\zeta_k+2\omega,
\end{equation}
\begin{equation}\label{25}
\zeta_k=\mu_k+W_1,\ \ \ \ (k=1,2),
\end{equation}
\begin{equation}\nonumber
\nu=E_2-E_1-2\omega,
\end{equation}
\begin{equation}\nonumber
\mu_{1,2}=\frac{1}{2}\biggr(\bar{\nu}\mp\sqrt{\bar{\nu}^2+4|\bar{\Omega}|^2}\biggl),
\end{equation}
\begin{equation}\nonumber
\bar{\nu}=\nu-(W_1-W_2)
\end{equation}
Quantities  $W_{\alpha}(\alpha=1,2)$ and $|\bar{\Omega}|$ are, respectively, the one-photon dynamic Stark shifts and the effective Rabi frequency of two-photon interaction:
\begin{equation}\label{26}
W_{\alpha}=\sum_k(\frac{|\Omega_{\alpha,k}|^2}{E_\alpha-E_k+\omega}+\frac{|\Omega_{\alpha,k}|^2}{E_\alpha-E_k-\omega}),\alpha=1,2
\end{equation}
\begin{equation}\label{27}
\bar{\Omega}=\sum_k \frac{\Omega^*_{1,k}\Omega^*_{k,2}}{E_1-E_k+\omega}
\end{equation}
where summation is performed over intermediate states.  	
We represent the solution to the Schrodinger equation in the form \eqref{24}, where $\varphi_k(t)$ (k=1,2) are the quasienergy wave functions  of two-photon resonant approximation.

In the manner similar to preceding Section we obtain for coefficients of the expansion (4) the system of equations of the form (5a,b), where we have, instead of \eqref{6},
\begin{equation}\label{28}
W^{(\alpha)}_k(E)=\biggl\{\begin{array}{c}
                      \beta_k U_E,\ \ \ \ \ \  \alpha=1 \\
                      \beta_k \Omega'_{E+\omega'}, \ \alpha=2
                    \end{array}
\end{equation}
From equation \eqref{5b} we now have
\begin{equation}\label{29}
b^{(\alpha)}_E(\lambda)=z_{\alpha}(\lambda)\delta (\lambda-E)+\frac{P}{\lambda-E}\sum_k W^{(\alpha)}_k(E)a_k(E).
\end{equation}
For determination of the first quasienergy wave function we choose
\begin{equation}\label{30}
z_{\alpha}(E)=z(E)\sum_k W^{(\alpha)}_k(E)a_k(E),
\end{equation}
then for $b^{(\alpha)}_E(\lambda)$ we have the expression \eqref{7}

By substituting the expression \eqref{7} into equations \eqref{5a} we obtain
\begin{equation}\label{31}
(\lambda -\lambda_k )a_k(\lambda)=\beta^{*}_k \biggr[z(\lambda)\biggr(|U(\lambda)|^2+|\Omega'_{\lambda+\omega'}|^2\biggl)+\Delta(\lambda)\biggl]X(\lambda),
\end{equation}
where
\begin{equation}\label{32}
\Delta (\lambda)=P\int \frac{|U_{\lambda}|^2+|\Omega'_{\lambda+\omega'}|^2}{\lambda-E}dE,
\end{equation}
\begin{equation}\label{33}
X(\lambda)=\sum_k \beta_k a_k(\lambda).
\end{equation}
After finding  $a_k(\lambda)$ from \eqref{31}, multiplying by $\beta_k$ and taking sum over $k$, we obtain for $X(\lambda)$ a homogenous algebraic equation:
\begin{equation}\label{34}
 \biggr\{1- \biggr[z(\lambda)\biggl(|U_{\lambda}|^2+|\Omega'_{\lambda+\omega'}|^2\biggl)+\Delta(\lambda)\biggl] \sum_k \frac{|\beta_k|^2}{\lambda-\lambda_k}\biggl\} X(\lambda)=0.
\end{equation}
Requiring existence of non-trivial solution to \eqref{34} gives for $z(\lambda)$ the following expression:
\begin{equation}\label{35}
z(\lambda)=\frac{2\pi}{\Gamma(\lambda)}\biggr[\frac{1}{\sum_k \frac{|\beta_k|^2}{\lambda-\lambda_k}}-\Delta(\lambda)\biggl],
\end{equation}
where $\Gamma(\lambda)$ is the full width of the resonance:
\begin{equation}\label{36}
\Gamma(\lambda)=2\pi\biggr(|U_{\lambda}|^2+|\Omega'_{\lambda+\omega'}|^2\biggl).
\end{equation}
After substituting $\beta_k$ ($k$=1,2) from \eqref{25} expression \eqref{35} for $z(\lambda)$ may be represented in the form
\begin{equation}\label{37}
z(\lambda)=\frac{2\pi}{\Gamma(\lambda)}\biggr[\frac{(\lambda-\lambda_1)(\lambda-\lambda_2)}{\lambda-\lambda_2+\mu_2}-\Delta(\lambda)\biggl].
\end{equation}
From the condition of orthonormalization of the first quasienergy wave function $\Phi^{(1)}_{\lambda}(t)$ (4):
\begin{equation}\label{38}
\langle\Phi^{(1)}_{\lambda'}(t)|\Phi^{(1)}_{\lambda}(t)\rangle=\delta(\lambda'-\lambda),
\end{equation}
we obtain for $X(\lambda)$
\begin{equation}\label{39}
X(\lambda)=\sqrt{\frac{2\pi}{\Gamma(\lambda)[z^2(\lambda)+\pi^2]}}.
\end{equation}
Finally, the first wave function appears to be
\begin{eqnarray}
\Phi^{(1)}_{\lambda} (t)=\sqrt{\frac{2\pi}{\Gamma(\lambda)[z^2(\lambda)+\pi^2]}}
\biggr\{\biggr[z(\lambda)\frac{\Gamma(\lambda)}{2\pi}+ \Delta (\lambda)\biggl] \nonumber\\
\sum_k e^{i\lambda_k t} \frac{\beta^*_k}{\lambda-\lambda_k}|\varphi_k(t)\rangle +\int\biggr[z(\lambda)\delta(\lambda-E)+\frac{P}{\lambda-E}\biggl]
\nonumber \\
\biggr[U_E|E\rangle_1+e^{-i\omega't}\Omega'_{E+\omega'}|E+\omega'\rangle_2\biggl]dE  \biggl\}.\label{40}
\end{eqnarray}

For determination of the second quasienergy wave function which should be orthogonal to the first one in \eqref{40} we require the following condition to be met \cite{21 U. Fano,28 L.C. Davis}:
\begin{equation}\label{41}
\sum_\alpha z_\alpha(\lambda)W^{(\alpha)*}_k(\lambda)=0;
\end{equation}
it follows that $a_j(\lambda)=0$ $j=1,2.$ From relationship \eqref{41} with allowance for \eqref{28} we have
\begin{equation}\label{42}
z_2(\lambda)=-\frac{U^*_\lambda}{\Omega'^*_{\lambda+\omega'}}z_1(\lambda);
\end{equation}
then
\begin{subequations}
\begin{eqnarray}
 \label{43a}   b^{(1)}_E(\lambda)=z_1(\lambda)\delta(\lambda-E),\\
 \label{43b}   b^{(2)}_E(\lambda)=-\frac{U^*_\lambda}{\Omega'^*_{\lambda+\omega'}}z_1(\lambda)\delta(\lambda-E).
\end{eqnarray}
\end{subequations}
Now by using (43a,b) in expression \eqref{4} for the second quasienergy wave function and the orthonormalisation condition
\begin{equation}\label{44}
    \langle\Phi^{(2)}_{\lambda'}(t)|\Phi^{(2)}_\lambda (t)\rangle=\delta(\lambda'-\lambda)
\end{equation}
we obtain
\begin{equation}\label{45}
    \Phi^{(2)}_\lambda(t)=\sqrt{\frac{2\pi}{\Gamma(\lambda)}}\biggr[\Omega'^*_{\lambda+\omega'}|\lambda\rangle_1+e^{-i\omega't}U_\lambda|\lambda+\omega'\rangle_2\biggl].
\end{equation}
It is easy to check directly that the functions $\Phi^{(1)}_\lambda (t)$ and $\Phi^{(2)}_{\lambda} (t)$ are orthogonal.
\section{Cross-sections of elastic and inelastic scattering; scattering length}
\subsection{Case of one-photon resonance}

Asymptotic ($r\rightarrow\infty$) expressions for continuous-spectrum wave functions with orbital angular momentum  $l$ are known to have the following appearance:
\begin{widetext}
\begin{subequations}
\begin{eqnarray}
\label{46a}|E\rangle^{(l)}_1\varpropto\frac{1}{k_1r_1}\sin{\biggr(k_1 r_1+\delta_l-\frac{1}{2}\pi l\biggl)}P_l(\cos{\Theta_1}), \\
\label{46b} |E+\omega'\rangle^{(l)}_1\varpropto\frac{1}{k_2 r_2}\sin{\biggr(k_2r_2+\delta_l-\frac{1}{2}\pi l\biggl)}P_l(\cos{(\Theta_2)})
\end{eqnarray}
\end{subequations}
\end{widetext}
with $P_l(\cos{\Theta})$ being the Legendre polynomials.

Taking into account that the wave functions of the bound states of atoms vanish asymptotically ($|\varphi_k(t)\rangle=0$ at $r\rightarrow\infty$) we can write the functions \eqref{8} after inserting (46a,b) as follows:
\begin{widetext}
\begin{eqnarray}\label{47}
\Phi^{(l)}_{\lambda,j}(t)\varpropto-\frac{\pi}{k_1(\lambda)r_1}X^{(l)}_{j,1}\frac{1}{\sin{\eta^{(j)}_l(\lambda)}} \sin{\biggr[k(\lambda)r_1+\delta_l+\eta^{(j)}_l(\lambda)-\frac{l\pi}{2}\biggl]}P_l(\cos{\Theta_1})- \nonumber\\ -\frac{\pi}{k_2(\lambda+\omega')r_2}X^{(l)}_{j,2}\frac{e^{-i\omega' t}}{\sin{\eta^{(j)}_l(\lambda)}}\sin{\biggr[k(\lambda+\omega')r_2+\delta_l+\eta^{(j)}_l(\lambda)-\frac{l\pi}{2}\biggl]}P_l(\cos{\Theta_2}),
\end{eqnarray}
\end{widetext}
where $\delta_l$ is the phase of potential (non-resonant) scattering and $\eta^{(j)}_l(\lambda)$ is the phase caused by resonant (Feshbach resonance) scattering; subscript j(j=1, 2) corresponds to two values of the function $z^{(j)}_l(\lambda)$ in \eqref{16}:
\begin{equation}\label{48}
\tan\eta^{(j)}_l(\lambda)=-\frac{1}{\bar{z}^{(j)}_l(\lambda)},\ \ \ \ \ \bar{z}^{(j)}_l(\lambda)=\frac{z^{(j)}_l(\lambda)}{\pi}.
\end{equation}
We now represent the scattering state vector at $r\rightarrow\infty$ as a superposition of quasienergy functions (47)
\begin{widetext}
\begin{eqnarray}\label{49}
\Phi_{\lambda}(t,1\rightarrow 1,2)=-\sum_j\sum_l B^{(l)}_j(\lambda)(2l+1)i^l e^{i(\delta_l+\eta^{(j)}_l(\lambda))}
\biggr\{\frac{X^{l}_{j1}(\lambda)}{k_1(\lambda)r_1}\sin{\biggr[k_1(\lambda)r_1+\delta_l+\eta^{(j)}_l(\lambda)-\frac{l\pi}{2}\biggl]}P_l\cos{(\Theta_1)}+\nonumber \\ +\frac{X^{(l)}_{j,2}}{k_2(\lambda+\omega')r_2}e^{-i\omega't}\sin{\biggr[k_2(\lambda+\omega')r_2+ \delta_l+\eta^{(j)}_l(\lambda)-\frac{l\pi}{2}\biggl]}P_l\cos{(\Theta_2)}\biggl\}.
\end{eqnarray}
\end{widetext}
Then, if we require the presence of incoming and outgoing waves in the first, elastic, channel and the absence of incoming wave in the second, inelastic, channel, we can write for the expansion coefficients in (49) $B^{(l)}_j(\lambda)$ the following relationship:
\begin{subequations}
\begin{eqnarray}
\label{50a} B^{(j)}_1(\lambda)=\frac{X^{(l)}_{2,2}}{\det[X^{(l)}_{i,j}(\lambda)]},\\
\label{50b} B^{(j)}_2(\lambda)=\frac{X^{(l)}_{1,2}}{\det[X^{(l)}_{i,j}(\lambda)]}.
\end{eqnarray}
\end{subequations}
Expressions for elastic and inelastic cross-sections have the forms:
\begin{subequations}
\begin{eqnarray}
\label{51a} \sigma_{el}=\frac{\pi}{k^2}\sum_l(2l+1)|1-S_l(\lambda)|^2,\\
\label{51b} \sigma_{inel}=\frac{\pi}{k^2}\sum_l(2l+1)(1-|S_l(\lambda)|^2).
\end{eqnarray}
\end{subequations}
Here $S_l(\lambda)$ is the diagonal element of the S-matrix,
\begin{equation}\label{52}
S_l(\lambda)=A^{(l)}_1(\lambda)S^{(1)}_l(\lambda)+A^{(l)}_2(\lambda)S^{(2)}_l(\lambda),
\end{equation}
where
\begin{equation}\label{53}
S^{(j)}_l(\lambda)=e^{2i(\delta_l+\eta^{(j)}_l(\lambda))}, (j=1,2),
\end{equation}
and factors $A^{(l)}_j(\lambda)$ $(j=1,2,3)$ are defined as
\begin{subequations}
\begin{eqnarray}
\label{54a} A^{(l)}_{1,2}(\lambda)=\frac{z^{(1,2)}_l(\lambda)}{z^{(1)}_l(\lambda)-z^{(2)}_l(\lambda)}(1-z^{(2,1)}_l(\lambda)C_{1,2}(\lambda)),\\
A^{(l)}_{3}(\lambda)=\frac{1}{z^{(1)}_l(\lambda)-z^{(2)}_l(\lambda)}\times\nonumber\\ \label{54b}\frac{(1-z^{(1)}_l(\lambda)C_1(\lambda))(1-z^{(2)}_l(\lambda)C_2(\lambda))}{C^*_{(1,2)}(\lambda)}.
\end{eqnarray}
\end{subequations}
Let us separate in (51a,b) the resonant state with orbital angular momentum $L$ and write the potential part of scattering cross-section in the form
\begin{equation}\label{55}
\sigma_{pot}=\frac{4\pi}{k^2}\sum_{l \neq L} (2l+1)\sin^2{\delta_l}.
\end{equation}
The total cross-section of elastic scattering we write as
\begin{equation}\label{56}
\sigma^{(L)}_{tot}=\sigma_{pot}+\sigma^{(L)el}_{res},
\end{equation}
where the resonance term is defined by the expression
\begin{widetext}
\begin{equation}\label{57}
\frac{\sigma^{(L)el}_{res}}{\sigma^{(L)}_{nonres}}=|A^{(L)}_1(\lambda)|^2F_1(\lambda)+|A^{(L)}_2(\lambda)|^2F_2(\lambda)+2|A^{(L)}_3(\lambda)|^2F_3(\lambda).
\end{equation}
\end{widetext}
Cross-section with orbital angular momentum $L$ in the absence of the resonance state \cite{29 H. Messy} is denoted $\sigma^L_{nonres}$ and equals
\begin{equation}\label{58}
\sigma^{(L)}_{nonres}=\frac{4\pi}{k^2}(2L+1)\sin^2{\delta_L}.
\end{equation}
The quantities $F_i(\lambda)$ ($i$=1,2,3) entering (57) are defined as
\begin{widetext}
\begin{subequations}
\begin{eqnarray}
\label{59a}
F_j(\lambda)=\frac{(\chi_L+\bar{z}^{(j)}_L(\lambda))^2}{1+\bar{z}^{(j)}_L(\lambda)^2} \ \ \ (j=1,2),\\
\label{59b}
F_3(\lambda)=\frac{\chi^2_L(1+\bar{z}^{(1)}_L(\lambda)\bar{z}^{(2)}_L(\lambda))(\chi_L+\bar{z}^{(1)}_L(\lambda))(\chi_L+\bar{z}^{(2)}_L(\lambda))}{(1+\bar{z}^{(1)}_L(\lambda))(1+\bar{z}^{(2)}_L(\lambda))},
\end{eqnarray}
\end{subequations}
\end{widetext}
where
\begin{equation}\label{60}
\chi_L=-\cot{\delta_L}.
\end{equation}
With use of expression (51b) we obtain for cross-section of inelastic resonant scattering with orbital angular momentum $L$:
\begin{equation}\label{61}
\sigma^{(L)}_{inel}=\frac{4\pi}{k^2}(2L+1)|A_3(\lambda)|^2\sin^2{(\eta^{(2)}_L(\lambda)-\eta^{(1)}_L(\lambda))}.
\end{equation}
We note that the obtained expressions for the cross-sections contain two scattering phases $\eta^{(1)}_L(\lambda)$ and $\eta^{(2)}_L(\lambda)$, which are connected by the relationship
\begin{equation}\label{62}
\tan{\eta^{(1)}_L(\lambda)}+\tan{\eta^{(2)}_L(\lambda)}=-\sum_\mu\frac{\frac{\tilde{\Gamma}^{(L)}_\mu(\lambda)}{2}}{\lambda- \tilde{\lambda}_{\mu}},
\end{equation}
where
\begin{eqnarray}
    \tilde{\Gamma}^{(\alpha,L)}_\mu(\lambda)=2\pi|\tilde{W}^{(\alpha,L)}_\mu(\lambda)|^2,\nonumber\\
\label{63}    \tilde{\Gamma}^{(L)}_\mu(\lambda)=\sum_{\alpha}\tilde{\Gamma}^{(\alpha,L)}_\mu(\lambda),\ \ \alpha,\mu=1,2.
\end{eqnarray}
Introducing a reduced value for quatities $\bar{z}^{(L)}_j(\lambda)$:
\begin{equation}\label{64}
    Z_{L}(\lambda)=\frac{\bar{z}^{(1)}_L(\lambda)\bar{z}^{(2)}_L(\lambda)}{\bar{z}^{(1)}_L(\lambda)+\bar{z}^{(2)}_L(\lambda)},
\end{equation}
expression (62) may be rewritten as
\begin{equation}\label{65}
    Z_{L}(\lambda)=\biggl[\sum_\mu\frac{\frac{\tilde{\Gamma}^{(L)}_\mu(\lambda)}{2}}{\lambda-\tilde{\lambda}_\mu}\biggr]^{-1}.
\end{equation}
If there is no inelastic channel ($\Omega'_E\rightarrow 0$) we pass to the case of purely elastic scattering. In this case we have $\bar{z}^{(1)}_L(\lambda)\rightarrow\infty, \  \bar{z}^{(2)}_L(\lambda)\rightarrow\bar{z}_{L}(\lambda),\  A^{(L)}_1(\lambda)\rightarrow 0,\  A^{(L)}_2(\lambda)\rightarrow 1, \ A^{(L)}_3(\lambda)\rightarrow 0.$ Here $\bar{z}_{L}(\lambda)$ has the form (65), where
\begin{eqnarray}\label{66}
    \tilde{\Gamma}^{(L)}_\mu(\lambda)=2\pi|\tilde{W}_\mu(\lambda)|^2,\nonumber\\
    \tilde{W}_\mu(\lambda)=\tilde{\alpha}_\mu(\lambda)\Omega_E+\tilde{\beta}_\mu(\lambda)U_E.
\end{eqnarray}

Expression (57) for cross-section of elastic scattering in the absence of inelastic channel may be represented in the form \cite{29 H. Messy}
\begin{equation}\label{67}
    \frac{\sigma^{(L)el}_{res}}{\sigma^{(L)}_{nonres}}=\frac{(\chi_L+\bar{z}_L(\lambda))^2}{1+z^2_L(\lambda)}.
\end{equation}
Cross-section of elastic resonant scattering in the absence of inelastic channel and in the presence of external magnetic field and laser radiation can be represented as
\begin{widetext}
\begin{equation}\label{68}
    \sigma^{(L)el}_{res}=4\pi\frac{2L+1}{k^2(\lambda)}\biggr|\frac{\frac{\tilde{\Gamma}_1}{2}(\lambda-\tilde{\lambda}_2) +\frac{\tilde{\Gamma}_2}{2}(\lambda-\tilde{\lambda}_1)} {(\lambda-\tilde{\lambda}_1)(\lambda-\tilde{\lambda}_2)+i\frac{\tilde{\Gamma}_1}{2}(\lambda-\tilde{\lambda}_2)+ i\frac{\tilde{\Gamma}_2}{2}(\lambda-\tilde{\lambda}_1) }\biggl|^2.
\end{equation}
\end{widetext}
Here $k(\lambda)=\sqrt{\frac{m\lambda}{\hbar^2}}$ with $m$ being the mass of an atom.

Depicted in Fig.3 are the cross-sections of resonant scattering in case of one-photon resonance between the laser radiation and the two molecular discrete states, at different values of  $\chi_L$. Figures 3(a,b) plot the cross-sections of elastic scattering in, respectively, absence and presence of inelastic channel, and Fig.3(c) the cross-section of inelastic scattering. As seen in the obtained plots, in elastic scattering (Fig.3a,b) the resonance is sharper at higher values of parameter $\chi_L$. With the decrease in this parameter the resonance becomes flatter and at $\chi_L=0$ we observe an antiresonance, because of interference between potential and resonant scattering.

\begin{figure}
  \includegraphics[width=7cm]{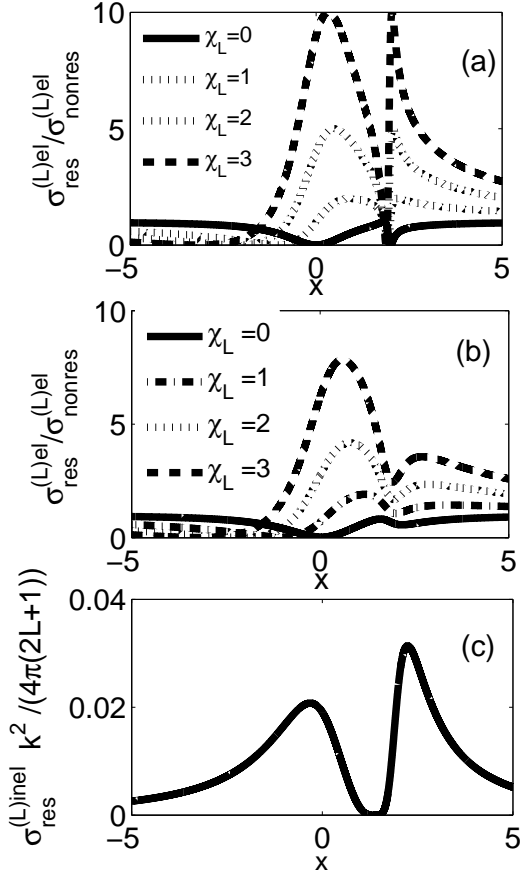}
  \caption{Plots of cross-sections of atom-atom collision in external magnetic and laser fields in the case of one photon resonance between the laser radiation and two molecular discrete states, versus the scaled energy $x=2(\lambda-\bar{\lambda}_1)/\tilde{\Gamma}^{(1)}_1$ for the values $\tilde{\Gamma}^{(1)}_2=0.2$,\ $\tilde{\Gamma}^{(2)}_1=0.4$, \ $\tilde{\Gamma}^{(2)}_2=0.8$, $\Omega=1$, $\nu=0$  (all quantities are in units of $\tilde{\Gamma}^{(1)}_1/2$): (a) and (b) cross-sections of elastic scattering in, respectively, the absence and presence of inelastic channel. The insets show the values of  $\chi_L$; (c) cross-section of inelastic scattering.}\label{Fig3}
\end{figure}
For determination of the scattering length we introduce a complex $\bar{\eta}_L(\lambda)$ from the relation \eqref{52}:
\begin{equation}\label{69}
    S_L(\lambda)=e^{2i\bar{\eta}_L(\lambda)}.
\end{equation}
This phase equals, according to \eqref{52}
\begin{equation}\label{70}
    \bar{\eta}_L(\lambda)=\delta_L+\frac{1}{2i}\ln\sum_j A^{(L)}_j(\lambda)e^{2i\eta^{(j)}_L(\lambda)}.
\end{equation}
From expression (48) we have
\begin{equation}\label{71}
    e^{2i\eta^{(L)}_j(\lambda)}=\frac{\bar{z}^{(j)}_L(\lambda)-i}{\bar{z}^{(j)}_L(\lambda)+i}
\end{equation}
and (70) takes the form:
\begin{equation}\label{72}
    \bar{\eta}_L(\lambda)=\delta_L+\frac{1}{2i}\ln\sum_jA^{(L)}_j(\lambda)\frac{\bar{z}^{(j)}_L(\lambda)-i}{\bar{z}^{(j)}_L(\lambda)+i}.
\end{equation}
Let us define the "effective" scattering length as
\begin{equation}\label{73}
    a^{(L)}_{eff}(k)=-\frac{\bar{\eta}_L(k)}{k};
\end{equation}
substitution of (72) into (73) provides
\begin{equation}\label{74}
    a^{(L)}_{eff}(k)=a^{(L)}_{0}(k)+a^{(L)res}_{eff}(k),
\end{equation}
where $a^{(L)}_0(k)=-\frac{\delta_L(k)}{k}$ is the background length caused by the nonresonant scattering; in this case the quantity
\begin{equation}\label{75}
    a^{(L)res}_{eff}(k)=-\frac{1}{2i}\frac{1}{k}\ln\sum_jA^{(L)}_l(k)\frac{\bar{z}^{(j)}_L(k)-i}{\bar{z}^{(j)}_L(k)+i}
\end{equation}
is complex.

In case of cold atoms ($k\rightarrow0$) and zero orbital angular momentum (S-wave) we can define the "reduced" width $\tilde{\gamma}_i$ ($\tilde{\Gamma}_i=2\tilde{\gamma}_ik$) for laser-induced decays. For the scattering length at $k\rightarrow0$ we obtain
\begin{equation}\label{76}
    a=a_0+a_{res}
\end{equation}
where $a_0$ is the background value and $a_{res}$ is the resonant part of the scattering length
\begin{equation}\label{77}
    a_{res}=-(\frac{\tilde{\gamma}^{(L)}_1}{\tilde{\lambda}_1}+\frac{\tilde{\gamma}^{(L)}_2}{\tilde{\lambda}_2}),
\end{equation}
where $\tilde{\lambda}_1$ and $\tilde{\lambda}_2$ are defined in \eqref{13}. The scattering length (77) coincides with the expression for the case of elastic scattering in absence of inelastic channel. Expression (77) shows that the scattering length depends on the laser radiation filed strength and on the magnetic field induction
\subsection{Case of two-photons resonance}
Asymptotic expression of the continuous-spectrum wave function with orbital angular momentum $l$ have the form (46a,b). Taking $|\varphi_{k}\rangle=0$ ($r\rightarrow\infty$) we obtain:
\begin{widetext}
\begin{subequations}
\begin{eqnarray}
\Phi^{(1)}_{\lambda,l}(t)\propto \sqrt{\frac{2\pi}{\Gamma_l(\lambda)}}\biggr\{\frac{U_\lambda}{k(\lambda)r_1}\sin{\biggr[k(\lambda)r_1+\delta_l+\eta_l(\lambda)-\frac{\pi l}{2}\biggl]}P_l(\cos\Theta_1)+ \nonumber \\
e^{-i\omega't}\frac{\Omega'_{\lambda+\omega'}}{k(\lambda+\omega')r_2} \sin{\biggr[k(\lambda+\omega')r_2+\delta_l+\eta_l(\lambda)-\frac{\pi l}{2}\biggl]}P_l(\cos{\Theta_2})\biggl\}, \label{78a} \\
\label{78b}
\Phi^{(2)}_{\lambda,l}(t)\propto \sqrt{\frac{2\pi}{\Gamma_l(\lambda)}}\biggr\{\frac{\Omega'^*_{\lambda+\omega'}} {k(\lambda)r_1}\sin{\biggr[k(\lambda)r_1+\delta_l-\frac{\pi l}{2}\biggl]}P_l(\cos\Theta_1)- \nonumber \\
e^{-i\omega't}\frac{U^*_{\lambda}}{k(\lambda+\omega')r_2} \sin{\biggr[k(\lambda+\omega')r_2+\delta_l-\frac{\pi l}{2}\biggl]}P_l(\cos{\Theta_2})\biggl\},
\end{eqnarray}
\end{subequations}
\end{widetext}
where $\delta_l$ is the phase of potential (non-resonant) scattering and $\eta_l(\lambda)$ is the phase caused by resonant scattering (Feshbach resonance) (48).

The scattering state vector we represent again as superposition of wave functions (78a,b) at $r\rightarrow 0$:
\begin{eqnarray}
\Phi_\lambda(1\rightarrow 1,2)=\sum_l[B^{(1)}_l(\lambda)(2l+1)i^le^{2i(\delta_l+\eta_l(\lambda))}\Phi^{(l)}_{\lambda,l}(t) \nonumber\\
+B^{(2)}_l(\lambda)(2l+1)i^le^{2i(\delta_l+\eta_l(\lambda))}\Phi^{(2)}_{\lambda,l}(t)] \nonumber \\ \label{79}
\end{eqnarray}

We require again the presence of incoming and outgoing waves in the first (elastic) channel and absence of the incoming wave in the second (inelastic) channel and obtain:
\begin{equation}\label{80}
    B^{(1)}_l(\lambda)=U^*_\lambda\sqrt{\frac{2\pi}{\Gamma_l(\lambda)}},\ \ \
    B^{(2)}_l(\lambda)=\Omega'_{\lambda+\omega'}\sqrt{\frac{2\pi}{\Gamma_l(\lambda)}},
\end{equation}
where $\Gamma_l(\lambda)$ is the full width of the resonance with orbital angular momentum $l$.

Finally the scattering state vector has the following appearance:
\begin{widetext}
\begin{eqnarray}\label{81}
\Phi_\lambda(1\rightarrow 1,2)=\sum_l(2l+1)i^l\biggr\{\frac{1}{k(\lambda)r_1}\biggr[\sin{\biggr(k(\lambda)r_1-\frac{\pi l}{2}\biggl)}-\nonumber\\
-\frac{1}{2i}\biggr(1-\biggr[\frac{\Gamma^{(2)}_l(\lambda)}{\Gamma_l(\lambda)}+\nonumber+\frac{\Gamma^{(1)}_l(\lambda)}{\Gamma_l(\lambda)} e^{2i\eta_l(\lambda)}\biggl] e^{2i\delta_l}\biggl)e^{i(k(\lambda)r_1-\frac{\pi l}{2})}\biggl]\biggl\}P_l(\cos{\Theta_1})-\nonumber\\
-e^{i\omega't}\frac{1}{k(\lambda+\omega')r_2}\frac{\pi}{2i}\frac{U^*_\lambda\Omega'_{\lambda+\omega'}}{\frac{\Gamma_l(\lambda)}{2}} \biggr[1-e^{2i\eta_l(\lambda)}\biggl]e^{i(k(\lambda+\omega')r_2-\frac{\pi l}{2})}e^{2i\delta_l}P_l(\cos{\Theta_2})\biggl\},
\end{eqnarray}
\end{widetext}
where $\Gamma_l(\lambda)$ is the full partial width:
\begin{subequations}
\begin{eqnarray}
\label{82a}
\Gamma_l(\lambda)=\Gamma^{(1)}_l(\lambda)+\Gamma^{(2)}_l(\lambda),\\
\label{82b}
\Gamma^{(1)}_l(\lambda)=2\pi|U_l(\lambda)|^2,\ \ \ \Gamma^{(2)}_l(\lambda)=2\pi|\Omega'_{\lambda+\omega'}|^2.
\end{eqnarray}
\end{subequations}
Cross-sections of elastic and inelastic scattering are given in the forms in (51a,b) where
\begin{equation}\label{83}
S_l(\lambda)=\biggl[1-\frac{\Gamma^{(1)}_l(\lambda)}{\Gamma_l(\lambda)}\biggl(1-e^{2i\eta_l(\lambda)}\biggr)\biggr]e^{2i\delta_l}.
\end{equation}
The total cross-section of elastic scattering we represent in the forms (55,56). Separate, as above the resonance with orbital angular momentum $L$, the cross-section of elastic resonant scattering is obtained:
\begin{equation}\label{84}
\sigma^{(L)el}_{res}=\sigma^{(L)el}_{nonres}\biggl|1+e^{i(\delta_L+\bar{\eta}_L(\lambda))}\frac{\sin{(\bar{\eta}_L(\lambda))}}{\sin{\delta_L(\lambda)}}\biggr|^2.
\end{equation}
Here we introduce the complex phase $\bar{\eta}_L(\lambda)$
\begin{equation}\label{85}
S_L(\lambda)=e^{2i(\delta_L+\bar{\eta}_L)}
\end{equation}
and $\sigma^{(L)el}_{nonres}$ is cross-section with orbital moment $L$ in absence of resonance state \eqref{58} \cite{29 H. Messy}.
With use of expression \eqref{48} the cross section \eqref{84} may be displayed as
\begin{equation}\label{86}
\sigma^{(L)el}_{res}=\sigma^{(L)el}_{nonres}\biggr|1+(\chi_L-1)\frac{\Gamma^{(1)}_L(\lambda)}{\Gamma_L()\lambda}\frac{1}{\bar{z}_L(\lambda)+i}\biggl|^2,
\end{equation}
where $\chi_L$ is defined in \eqref{60}

Neglecting non-resonant terms we obtain for cross-sections:
\begin{widetext}
\begin{subequations}\label{87}
\begin{eqnarray}
\label{87a}
\sigma^{(L)el}_{res}=\frac{2\pi}{k^2}(2L+1) \frac{(\Gamma^{(1)}_L(\lambda))^2(\lambda-E_1-W_1-2\omega)^2}{|(\lambda-\bar{\lambda}_{-})(\lambda-\bar{\lambda}_{+})|^2},\\
\label{87b}
\sigma^{(L)inel}_{res}=\frac{2\pi}{k^2}(2L+1)
\frac{\Gamma^{(1)}_L(\lambda)\Gamma^{(2)}_L(\lambda)(\lambda-E_1-W_1-2\omega)^2}{|(\lambda-\bar{\lambda}_{-})(\lambda-\bar{\lambda}_{+})|^2}.
\end{eqnarray}
\end{subequations}
\end{widetext}

It follows from expressions for cross-sections of resonant scattering (87a,b)  that quenching of the resonance occurs at the energy $\lambda=E_1+W_1+2\omega$, that is at the energy of the lower stable level including its Stark shift plus the energy of two photons. Here $\bar{\nu}$ and $\bar{\Omega}$ are determined by formulas (25-27).

Cross-sections (87a,b) may also be represented as
\begin{subequations}
\begin{eqnarray}
\sigma^{(L)el}_{res}=\frac{2\pi}{k^2}(2L+1)\frac{(\Gamma^{(1)}_L)^2}{|R_+-R_-|^2}\biggr|\frac{R_+}{\lambda-\bar{\lambda}_+}-\frac{R_-}{\lambda-\bar{\lambda}_-}\biggl|^2,
\ \ \label{89a}\\
\sigma^{(L)inel}_{res}=\frac{2\pi}{k^2}(2L+1)\frac{\Gamma^{(1)}_L\Gamma^{(2)}_L}{|R_+-R_-|^2}\biggr|\frac{R_+}{\lambda-\bar{\lambda}_+}-\frac{R_-}{\lambda-\bar{\lambda}_-}\biggl|^2,
\ \ \label{89b}
\end{eqnarray}
\end{subequations}
where
\begin{subequations}
\begin{eqnarray}
\bar{\lambda}_{\mp}=E_1+W_1+2\omega+R_{\mp}, \ \ \label{88a} \\
R_{\mp}=\frac{1}{2}[\bar{\nu}+\Delta-i\frac{\Gamma_L}{2}\mp \nonumber \\ \mp\sqrt{(\bar{\nu}+\Delta-i\frac{\Gamma_L}{2})^2+4|\bar{\Omega}|^2}].
\label{88b}
\end{eqnarray}
\end{subequations}
\begin{figure}
  \includegraphics[width=8cm]{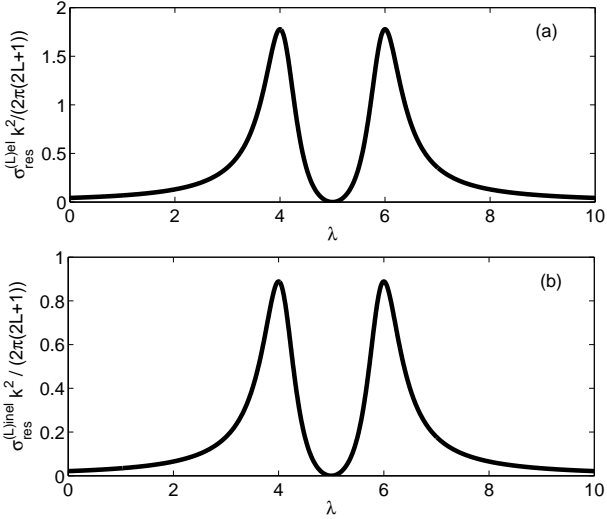}\\
  \caption{Plots of cross-sections of atom-atom collision in external magnetic and laser fields in the case of two-photon resonance between the laser radiation and two molecular discrete states, versus energy $\lambda$, for the values $E_1+W_1+2\omega=5,\bar{\nu}+\Delta=0,\Gamma_L=1.5,\bar{\Omega}=1$ (all quantities are in units of $\Gamma_L^{(1)})$. (a) elastic scattering, (b) inelastic scattering}\label{Fig4}
\end{figure}
Figure 4 demonstrates the cross-sections of elastic (a) and inelastic (b) resonance scattering in case of two-photon resonance between the laser radiation and two molecular discrete states.
Appearance of two peaks in cross-sections (88a,b) of resonant scattering is caused by the Autler-Towns effect.

In order to determine the scattering length, we use expression (85) to obtain
\begin{equation}\label{90}
    \bar{\eta}^{res}_L(\lambda)=\frac{1}{2i}\ln{\biggr [1-\frac{\Gamma^{(1)}_L(\lambda)}{\Gamma_L(\lambda)}(1-e^{2i\eta_L(\lambda)})\biggl]^{1/2}}.
\end{equation}
As in preceding sections, we define the �effective� scattering length (74).

In case of cold atoms and S-wave scattering it is again possible to define the �reduced� width $\gamma_i$ ($\Gamma_i=2\gamma_ik$) which is independent of k.
For the scattering length we obtain \eqref{76}, and
\begin{equation}\label{91}
    a_{res}=-\frac{1}{2i}\lim_{k\rightarrow 0}\biggr\{\frac{1}{k}\ln{\bigr[1-2i\frac{\Gamma^{(1)}_0(k)}{\Gamma_0(k)}\frac{1}{\bar{z}_0(k)+i}\biggl]}\biggl\}.
\end{equation}
Finally, the resonant scattering length appears as
\begin{equation}\label{92}
a_{res}=-\frac{\lambda_2-\mu_2}{\lambda_1\lambda_2}\gamma^{(1)}_0,
\end{equation}
where $\lambda_1$,$\lambda_2$ and $\mu_2$ are defined in (25).

We considered formation of Feshbach resonances with allowance for laser-induced inelastic collision at two-photon resonance between the lower vibrational molecular state in the open channel and higher-lying molecular vibrational state in the closed channel. In this case laser-induced decay does not occur and Feshbach resonances arise under influence of external magnetic field because of formation of dressed states via two-photon-resonant Autler-Townes mechanism. In this case it follows from the obtained expressions for cross-sections of elastic and inelastic scattering that full quenching of resonances takes place at the energy equal to that of the initial state, i.e., $\lambda=E_1+W_1+2\omega$ (energy of the lower stable state $E_1$ including the Stark shift $W_1$ and the energy of two photons). Such phenomenon is observed in, e.g., nuclear reactions and in resonant effects in elastic media.

At sufficiently high intensities of resonant laser radiation an asymmetry exists in dependence on the sign of two-photon detuning (effect of self-induced adiabatic passage of resonance) which is observed in the decay of system. However, similar effects are not observed in case of resonant scattering because of adiabatically turned interaction. It follows from the obtained expression (88a,b) that the real parts of quasienergies are crossing ($Re\lambda_-=Re\lambda_+$) when
\begin{equation}\label{93}
    Re\biggr[\biggr(\bar{\nu}+\Delta-i\frac{\Gamma_L}{2}\biggl)^2+4|\bar{\Omega}|^2\biggl]=0
\end{equation}
For meeting this condition it is necessary that $\bar{\nu}+\triangle=0$ and $2|\bar{\Omega}|\leq\frac{\Gamma_L}{2}$, while under condition  $2|\bar{\Omega}|>\frac{\Gamma_L}{2}$ anticrossing takes place. Hence, in the point $2|\bar{\Omega}|=\frac{\Gamma_L}{2}$  the interaction pattern does change essentially. At $2|\bar{\Omega}|=\frac{\Gamma_L}{2}$  and $\bar{\nu}+\Delta=0$ the poles of the resonance coincide ($\lambda_-=\lambda_+$) and we have a double-pole resonance. For revealing the double-peak in cross-sections of resonant scattering the condition
\begin{equation}\label{94}
|Re\lambda_--Re\lambda_+|>|Im\lambda_{\pm}|
\end{equation}
should be met. It then follows that at $\bar{\nu}+\Delta=0$  splitting effects are possible if the condition $2|\bar{\Omega}|>\frac{\Gamma_L}{2}$ is met. Under condition $2|\bar{\Omega}|\ll\frac{\Gamma_L}{2}$  we obtain from the expressions for cross-sections the well-known expressions for resonant scattering with an isolated level.
\section{\protect\normalsize Conclusion}

We considered collision of two atoms with formation of Feshbach resonance in the field of laser radiation. The interaction $U_E$ (notation in the Fig.1 and Fig.2) between electronic states means such an interaction with the external magnetic field where the bound state in the closed channel and the states of colliding atoms in the open channel are different hyperfine states. In the absence of this interaction, the quantity $U_E$ describes the weak interaction between the nuclear and electronic motions. Taking into account these terms lead, by employing the theory of perturbation, to arising of transitions between different electronic states.

In this work, we examined the cases where laser radiation couples the lower molecular vibrational state in the open channel with higher-lying molecular state in the closed channel, by one-photon resonance interaction, and with atomic states in continuum, by interaction  $\Omega_E$. Resonant laser radiation forms, via Autler-Townes effect, dressed states which produce, by interaction with the magnetic field, the Feshbach resonance. The interaction $\Omega_E$ which couples low-lying vibrational state with atomic continuum produces, via Autler-Townes effect and LICS, the laser-induced Feshbach resonances. Thus, we showed that combined effect of formation of Feshbach resonances takes place under action of both magnetic field and laser radiation. This enables control of Feshbach resonances by tuning both the magnetic field strength and the laser radiation frequency and power. In addition, the widths of resonances also depend on the magnetic field and laser intensity. The latter broadens the Feshbach resonance facilitating considerably the experimental observation of asymmetry in the total scattering cross-section arising from interference between the resonance and potential scattering. As seen in the obtained plots for elastic scattering (Fig.3a,b), the resonance is sharper at higher values of the parameter $\chi_L$. With the decrease in this parameter the resonance becomes flatter and at $\chi_L=0$ we observe an antiresonance, because of the interference between potential and resonant scattering. We also studied the influence of laser-induced inelastic decay channel on the considered phenomena.
We considered arising of Feshbach resonance with taking into account laser-induced inelastic collision in the case of two-photon resonance between the lower vibrational molecular state in open channel and higher-lying vibrational molecular state in closed channel. In this case, we obtained that the laser-induced decay of the lower vibrational state into continuum is absent and the Feshbach resonances are formed under action of the external magnetic field at formation of two-photon-resonance-dressed states via Autler - Townes effect. In this case, as follows from the expressions for cross-sections of elastic and inelastic scattering that we obtained in this work, complete quenching of resonances occurs at the energy equal to that of the initial state of considered system, i.e., $\lambda=E_1+W_1+2\omega$  (energy $E_1$ of the stable lower state with its Stark shift $W_1$ plus the energy $2\omega$  of two photons). Such phenomenon is observed in, e.g., nuclear reactions and in resonant effects in elastic media. By analyzing the expressions for quasienergies and poles of resonances, we showed existence of some asymmetry in the dependence on the magnetic field value and the laser field strength. We obtained that under condition $\nu+\Delta =0$ crossing of real parts of quasienergies takes place if $2|\Omega|\leq\Gamma_L/2$ and anticrossing if $2|\Omega|>\Gamma_L/2$. Hence, the interaction pattern does change essentially in the point $2|\Omega|=\Gamma_L/2$. In this case the poles of resonance coincide, $\lambda_-=\lambda_+$, and a double-pole resonance occurs. In order to observe the double peak in the cross-section of resonant scattering at $\nu+\Delta=0$, the condition $2|\Omega|>\Gamma_L/2$  should be met. Under condition $2|\Omega|\ll\Gamma_L/2$  we arrive at the well-known expressions for elastic and inelastic scattering on an isolated level.
It should be noted that the resonant peak dominates in the scattering cross-section at the resonance energy, which enables determination of the quantum number of angular momentum in the resonance state.
In all considered cases, we obtained expressions for the scattering length depending on the magnetic field value and laser radiation intensity.
The results obtained in this work may be employed for novel studies of collisions of cold atoms, in particular, of alkali metal atoms, such as $^{85,87}$Rb,$^{6,7}$Li, $^{133}$Cs, and so on, and performing new experiments in BECs.
\bigskip
\subsection*{Acknowledgment}

We are very grateful to Professor M.V. Fedorov for fruitful discussions.

Work was in part supported by the Ministry of Education and Science of Armenia (MESA), grant no. 11-1c124 of State Science Committee of MESA, IRMAS International Associated Laboratory,ANSEF-optPS-2911 and the Volkswagen Stiftung I/84 953.

\bigskip
\appendix*

\end{document}